\documentclass[a4paper,12pt]{iopart}
\usepackage{iopams}
\usepackage[dvips]{graphicx}
\usepackage{longtable}
\usepackage{color}
\newcommand {\nc} {\newcommand}
\nc {\p} {$+$}
\nc {\m} {$-$}
\nc {\s} {\,\,\,}
\nc {\sm} {\,\,\,\,\,\,\,}
\nc {\sn} {\,\,\,\,\,\,\,\,}
\nc {\beq} {\begin{eqnarray}}
\nc {\eeq} {\end{eqnarray}}
\nc {\eeqn} [1] {\label{#1} \end{eqnarray}}
\nc {\eoln} [1] {\label{#1} \\}
\nc {\eol} {\nonumber \\}
\nc {\la} {\mbox{$\langle$}}
\nc {\ra} {\mbox{$\rangle$}}
\nc {\ve} [1] {\mbox{\boldmath $#1$}}
\begin{document}
\title{Dipole transitions in the bound rotational-vibrational spectrum 
of the heteronuclear molecular ion HD$^+$}
\author{Horacio Olivares~Pil\'on\footnote{Present address: 
Instituto de Ciencias Nucleares, Universidad Nacional Aut\'onoma de
M\'exico, Apartado Postal 70-543, 04510 M\'exico, DF, Mexico} 
and Daniel Baye}
\address{Physique Quantique C.P. 165/82, and 
Physique Nucl\'eaire Th\'eorique et Physique Math\'ematique, C.P.\ 229, 
Universit\'e Libre de Bruxelles (ULB), B 1050 Brussels, Belgium}
\eads{\mailto{dbaye@ulb.ac.be, horop@nucleares.unam.mx}}
\begin{abstract}
The non-relativistic three-body Schr\"odinger equation of the heteronuclear molecular ion HD$^+$  
is solved in perimetric coordinates using the Lagrange-mesh method. 
Energies and wave functions of the four lowest vibrational bound or quasibound states $v=0-3$ 
are calculated for total orbital momenta from 0 to 47. 
Energies are given with an accuracy from about 12 digits for the lowest vibrational level 
to at least 9 digits for the third vibrational excited level. 
With a simple calculation using the corresponding wave functions, 
accurate dipole transition probabilities per time unit between those levels 
are given over the whole $v=0-3$ rotational bands. 
Results are presented with six significant figures. 
\end{abstract}
\submitto{\jpb}
\centerline{\today}
\maketitle
\section{Introduction}
\label{sec:intro}
The heteronuclear diatomic molecule HD$^+$ is the lightest isotopomer of H$_2^+$. 
The electronic ground state supports 637 rotational-vibrational levels 
of which 563 are bound and 74 are quasibound levels \cite{Mo93HD}. 
As for any three-body system, exact solutions of the Schr\"odinger equation 
can not be obtained but it is possible to reach a high accuracy for both energies and wave functions. 

Theoretical dissociation energies have been investigated for a long time. 
In 1993, Moss presented a detailed study including radiative and relativistic corrections 
of most of the bound and quasibound levels of the electronic ground state~\cite{Mo93HD} 
(see also several references to previous works). 
The dissociation energies are presented there in cm$^{-1}$ with nine figures. 
They correspond to total energies with about ten significant digits. 
Improvements in accuracy have been reached in several papers 
but for limited numbers of energy levels 
\cite{Mo99,HBG00,F02,YZL03,YZ04,K06,KH06,HNN09,TTZY12}. 
In particular, the ground state has been widely studied, but not always with the same mass values. 
The CODATA 1986 masses are employed in \cite{Mo93HD,Mo99,HBG00,F02,YZL03,YZ04} 
while the CODATA 2002 and 2006 values are used in \cite{K06,KH06} and \cite{HNN09,TTZY12}, 
respectively. 
To date, the most accurate result is determined with a 25-digit accuracy 
by Hijikata \etal \cite{HNN09}. 

The molecular ion HD$^+$ as well as the other heteronuclear isotopomers HT$^+$ and DT$^+$ 
have a permanent electric dipole moment. 
Hence, electric dipole (E1) transitions are possible. 
In 1953, the original work of Bates and Poots~\cite{BP53} included the theoretical Einstein A coefficients 
between some of the lowest vibrational-rotational levels. 
The lifetimes of 22 vibrational levels for the total orbital angular momentum $L=0$ 
were determined in \cite{PHB79}. 
Recently, Tian \etal \cite{TTZY12} published oscillator strengths 
for the six first rotational and six first vibrational levels. 

In this work, we extend to HD$^+$ our previous studies of transition probabilities 
within the bound spectra of H$_2^+$ \cite{OB12T} and D$_2^+$ \cite{OP13D2}. 
The calculations are simplified by the fact that E1 transitions are not forbidden 
like in H$_2^+$ and D$_2^+$ but they are made heavier 
by the lower symmetry of the three-body wave functions. 
Accurate energies are calculated beyond the Born-Oppenheimer approximation 
with the Lagrange-mesh method in perimetric coordinates \cite{HB99,HB01,HB03} 
with which the calculation is particularly simple and very precise. 
E1 transition probabilities are obtained from the corresponding three-body wave functions. 

The Lagrange-mesh method is an approximate variational calculation using a basis of Lagrange 
functions and the associated Gauss quadrature. 
It has the high accuracy of a variational approximation and the simplicity of a calculation 
on a mesh. 
The most striking property of the Lagrange-mesh method is that, in spite of its simplicity, 
the obtained energies and wave functions can be as accurate with the Gauss quadrature 
approximation as in the original variational method with an exact calculation 
of the matrix elements \cite{BHV02,Ba06}. 
 
The Lagrange-mesh method also provides analytical approximations for the wave functions 
that lead to very simple expressions for a number of matrix elements when used with the 
corresponding Gauss-Laguerre quadrature. 
In the same direction, this method has been applied to obtain not only energies 
and quadrupole transitions probabilities of the homonuclear systems H$_2^+$~\cite{OB12T} 
and D$_2^+$~\cite{OP13D2} but also polarizabilities for H$_2^+$ \cite{OB12P}. 
We keep here the CODATA1986 masses to allow a full comparison with the complete 
set of energies of \cite{Mo93HD}. 
The transition probabilities are presented with six significant figures 
and do not change with later mass conventions. 

In section \ref{sec:lmtp}, the expressions for the transition probabilities are summarized.
Lagrange-mesh expressions for the transition matrix elements are presented briefly. 
In section \ref{sec:res}, energies are given for the lowest four vibrational 
levels over the full rotational bands and E1 transition probabilities are tabulated. 
Concluding remarks are presented in section \ref{sec:conc}. 
\section{Lagrange-mesh calculation of transition probabilities}
\label{sec:lmtp}
In the three-body Hamiltonian that we are considering, 
two nuclei with masses $m_1$ and $m_2 < m_1$ and charges $q_1$ and $q_2$ 
and an electron with mass $m_e=1$ and charge $q_e=-1$ (in atomic units) 
interact through Coulomb forces. 
The total orbital momentum $L$ and parity $\pi$ are constants of motion. 
The dimensionless dipole oscillator strength for an electric transition  
between an initial state $i$ with energy $E_i$ and a final state $f$ 
with energy $E_f$ is given by \cite{So72,So79}
\beq
f_{i\rightarrow f}^{(1)} = \frac{2}{3} (E_i - E_f) 
\frac{|\la i L_i || d^{(1)} || f L_f \ra |^2}{2L_i+1}\,,
\eeqn{2.1}
where $d^{(1)}_\mu$ is the electric dipole operator defined below. 
The transition probability per time unit for $E_i > E_f$ is given in atomic units by 
\beq
W_{i\rightarrow f}^{(1)} = 2\alpha^3(E_i - E_f)^{2}f_{i\rightarrow f}^{(1)},
\eeqn{2.2}
where $\alpha$ is the fine-structure constant. 
Lifetimes can be calculated using 
\beq
\tau = \Bigg( \sum_{E_f < E_i} W_{i \rightarrow f}^{(1)} \Bigg)^{-1}.
\eeqn{2.3}

In the center-of-mass frame, the Hamiltonian 
is written as a function of two pseudo-Jacobi coordinates 
\beq
\begin{array}{rcl}
\ve{R} &=&  \ve{r}_{e1}-\ve{r}_{e2}, \\
\ve{r} &=& (\ve{r}_{e1}+\ve{r}_{e2})/2,
\end{array}
\eeqn{3.1}
where $\ve{r}_{ei}$ is the coordinate of the electron with respect to nucleus $i$. 
These coordinates are expressed as a function of three Euler angles 
$\psi, \theta, \phi$ and three internal perimetric coordinates $x,y,z$ 
which are defined as linear combinations of the interparticle distances \cite{Pe58}, 
\beq
\begin{array}{rcl}
x &=& R-r_{e2}+r_{e1}, \\
y &=& R+r_{e2}-r_{e1}, \\
z &=& -R+r_{e2}+r_{e1}.
\end{array}
\eeqn{3.2} 
The domains of variation of these six variables are $[0,2\pi]$ for $\psi$ and $\phi$, $[0,\pi]$
for $\theta$ and $[0,\infty[$ for $x, y$ and $z$. 

In perimetric coordinates, the dipole tensor operator reads 
\beq
d^{(1)}_{\mu}={\mathcal A}^{(1)}_0 D^{1}_{\mu 0} 
+ {\mathcal A}^{(1)}_1\left(D^{1}_{\mu 1}-D^{1}_{\mu -1} \right) \,,
\eeqn{3.3}
where 
\beq
{\mathcal A}^{(1)}_0 = (q'_2-q'_1) \frac{R}{2} + q'_e \zeta,
\quad 
{\mathcal A}^{(1)}_1 = q'_e \frac{\rho}{\sqrt{2}}
\eeqn{3.5}
with
\beq
q'_i = q_i-\frac{Q}{M}m_i, \quad i=1,\, 2,\, e. 
\eeqn{3.6}
In these expressions, $M$ and $Q$ are the total mass and charge of the system, 
respectively, and $D^{1}_{\mu q}(\psi,\theta,\phi)$ is a Wigner matrix element. 
For $m_1 = m_2$ and $q_1 = q_2$, one recovers equation (10) of \cite{OB12P}. 
The definitions and expressions of $R$, $\rho$ and $\zeta$ in terms 
of the perimetric coordinates are given for example in \cite{HB99,HB01}. 
The dipole operator is odd under space reflections. 

The wave functions with orbital momentum $L$ and parity $\pi$ are expanded as \cite{HB03} 
\beq
\Psi^{(L^{\pi})}_M = \sum_{K=0}^L {\cal D}_{MK}^{L\pi}(\psi,\theta,\phi) \Phi_K^{(L^{\pi})}(x,y,z). 
\eeqn{4.1}
where the ${\cal D}_{MK}^{L\pi}(\psi,\theta,\phi)$ are normalized parity-projected Wigner functions 
\cite{HB03,OB12T}. 
In practice, the sum can be truncated at some value $K_{\rm max}$. 

Due to the permanent dipole moment of the molecule, E1 transitions are possible 
within the $\Sigma_g$ rovibrational band, between states of different parities. 
For transitions between natural-parity states $\pi_{i,f} = (-1)^{L_{i,f}}$, 
the dipole strength is given by 
\beq 
|\la i L_i || d^{(1)} || f L_f \ra |^2 
= (2L_i+1) \bigg| \sum_{\kappa=0}^{1} 
\sum_{K_i K_f}  F_{K_iK_f\kappa}^{L_iL_f1} A_{K_i K_f;\kappa}^{L_i L_f;1} \bigg|^2,
\eeqn{4.8}
where $L_f = |L_i \pm 1|$ and 
\begin{eqnarray} 
F_{K_iK_f0}^{L_iL_f1} = (L_i1K_i0|L_fK_f),
\eol \fl
F_{K_iK_f1}^{L_iL_f1} = (L_i1K_i1|L_fK_f)(1+\delta_{K_i0})^{1/2}-(L_i1K_i-1|L_fK_f)(1+\delta_{K_f0})^{1/2}.
\label{eq:15}
\end{eqnarray}
The matrix elements 
\beq
A_{K_i K_f;\kappa}^{L_i L_f;1} = \la \Phi_{K_f}^{f(L_f^{\pi_f})} | 
\mathcal{A}^{(1)}_\kappa | \Phi_{K_i}^{i(L_i^{\pi_i})} \ra
\eeqn{4.6}
are calculated by integration over the perimetric coordinates with the volume element 
$(x+y)(y+z)(z+x)dxdydz$. 

The $\Phi^{(L^{\pi})}_K(x,y,z)$ functions of equation \eref{4.1} are expanded 
in the Lagrange basis as 
\beq
\Phi^{(L^{\pi})}_K(x,y,z) &=& \sum_{i=1}^{N_x} \sum_{j=1}^{N_y} \sum_{k=1}^{N_z} 
C_{Kijk}^{(L^{\pi})} F^K_{ijk}(x,y,z).
\eeqn{5.1}
The three-dimensional Lagrange functions $F^K_{ijk}(x,y,z)$ are infinitely 
differentiable functions satisfying the Lagrange property 
with respect to the three-dimensional mesh $(h_x u_p,h_y v_q,h_z w_r)$, 
\beq
F^K_{ijk}(h_x u_p,h_y v_q,h_z w_r)\propto \delta_{ip}\delta_{jq}\delta_{kr},
\eeqn{5.2}
i.e.\ they vanish at all mesh points but one. 
The mesh points $(h_x u_p,h_y v_q,h_z w_r)$ correspond to the zeros $u_p$, $v_q$, $w_r$ 
of Laguerre polynomials of respective degrees $N_x$, $N_y$, $N_z$. 
Three scale parameters $h_x, h_y, h_z$ are introduced in order to fit 
the mesh to the size of the actual physical problem. 
See Refs.~\cite{HB99,HB01,HB03,OB12T,OB12P} for details. 

The three-body Hamiltonian in perimetric coordinates for each good quantum number $L$ 
and its discretization on a Lagrange mesh are given in \cite{HB01}. 
The eigenvalues and eigenvectors of this large sparse matrix 
are calculated with the package JADAMILU \cite{BN07}. 
For given $L^\pi$, the eigenvalues in increasing order are labeled by the quantum number 
$v \ge 0$ related to the vibrational excitation in the Born-Oppenheimer picture. 
The corresponding eigenvectors provide the coefficients appearing in expansion \eref{5.1}. 

Let us consider initial and final components \eref{5.1} with respective coefficients 
$C_{K_iijk}^{i(L_i^{\pi_i})}$ and $C_{K_fijk}^{f(L_f^{\pi_f})}$. 
Because of the Lagrange property \eref{5.2}, 
the matrix elements \eref{4.6} are simply given by \cite{OB12P} 
\beq
A_{K_i K_f;\kappa}^{L_i L_f;1} 
\approx \sum_{i=1}^{N_x} \sum_{j=1}^{N_y} \sum_{k=1}^{N_z} 
C_{K_iijk}^{i(L_i^{\pi_i})} C_{K_fijk}^{f(L_f^{\pi_f})} 
\mathcal{A}^{(1)}_\kappa(h u_i,h v_j,h_z w_k)
\eeqn{5.3}
by using the Gauss quadrature associated with the mesh. 
\section{Energies and E1 transition probabilities}
\label{sec:res}
\subsection{{\rm HD}$^+$ rovibrational spectrum}
Energies of natutal-parity states are calculated 
for the four lowest vibrational levels $v = 0-3$ 
of the total orbital angular momenta  $L=0-47$. 
A high accuracy is obtained for most states by choosing $N_x = N_y = 40$, 
$N_z = 14$ and $h_x = h_y = 0.11$, $h_z = 0.4$. 
We did not find any significant advantage in choosing $N_x \ne N_y$ or $h_x \ne h_y$. 
The total number of basis states is 22400 for each value of $K$. 
In most calculations ($L \ge 2$), we choose $K_{\rm max} = 2$. 
Then the size of the matrix is 67200. 

For the sake of comparison with energies of other works 
and specially with the very complete results of Moss \cite{Mo93HD}, 
the 1986 fundamental constants are used everywhere, 
$m_p=1836.152\,701$ and $m_d=3670.483\,014$ in atomic units, 
except in Tables \ref{tab:2} and \ref{tab:6}. 
The exceptions occur for the comparison with the oscillator strengths of \cite{TTZY12} 
where the more recent 2006 values are used. 

Because of the different masses of the nuclei, two two-body dissociations are possible 
with thresholds $E_d^{{\rm\, H+D}^+} = -0.499\,727\,839\,716$ 
and $E_d^{{\rm\, D+H}^+} = -0.499\,863\,815\,249$ a.u. 
The small gap between these two thresholds does not make any quantitative difference 
in the number of bound states. 
In Table \ref{tab:1}, the obtained energies are presented as the first line for each $L$ value. 
The accuracy is estimated from the stability of the digits with respect to test calculations 
with smaller and larger numbers of mesh points (see Table \ref{tab:3} below). 
The error is expected to be at most of a few units on the last displayed digit. 
Literature results, sometimes truncated and rounded, are displayed in the following lines. 
Except in the low-$L$ or low-$v$ regions where results of other references are mentioned, 
the literature results are the 10-digit energies obtained by Moss in \cite{Mo93HD}. 
Since the energies of \cite{Mo93HD} are converted here from cm$^{-1}$ into atomic units, 
their accuracy is about 2-3 units on the last digit.

\begin{center}
\begin{longtable}{rllll}
\caption{Energies of the four lowest vibrational bound or quasibound levels 
in the $\Sigma_g$ rotational band of the HD$^+$ molecular ion. 
Quasibound levels are separated from bound levels by a horizontal bar. 
For each $L$ value, the Lagrange-mesh energies obtained with $N_x = N_y = 40$, $N_z = 14$ 
and $h_x = h_y = 0.11$, $h_z = 0.4$ are presented in the first line. 
The results of Moss \cite{Mo93HD} are indicated without superscript in the next lines. 
More accurate energies are given for some levels 
(a:~\cite{F02}, b:~\cite{YZL03}, c:~\cite{HBG00}, d:~\cite{Mo99}, e:~\cite{YZ04}). 
The proton and deuteron mass are taken as $m_p=1836.152\,701$ 
and $m_d=3670.483\,014$, respectively.}
\label{tab:1}\\
\hline
$L$& $v=0$          & $v=1$           &  $v=2$          &$v=3$\\
\hline
\endfirsthead
\multicolumn{5}{c}{{\tablename} \thetable{} -- Continuation}\\ 
\hline
$L$& $v=0$          & $v=1$           &  $v=2$          &$v=3$\\
\hline
\endhead
\hline
\multicolumn{5}{l}{{Continued on Next Page\ldots}}\\
\endfoot
\hline
\endlastfoot
 0&-0.597\,897\,968\,645\,1&-0.589\,181\,829\,652&-0.580\,903\,700\,33&-0.573\,050\,546\,1\\ 
  &-0.597\,897\,968\,645\,036$^{a,b}$&&&\\
  &-0.597\,897\,968\,644\,84$^{c}$&-0.589\,181\,829\,653\,3$^{c}$
  &-0.580\,903\,700\,369$^{c}$    &-0.573\,050\,546\,8$^{c}$\\
  &-0.597\,897\,968\,645\,0$^d$&-0.589\,181\,829\,653\,8$^d$&-0.580\,903\,700\,6 &-0.573\,050\,546\,9\\ 
 1&-0.597\,698\,128\,231\,2&-0.588\,991\,112\,090&-0.580\,721\,828\,23&-0.572\,877\,276\,7\\ 
  &-0.597\,698\,128\,231\,122$^a$&&&\\
  &-0.597\,698\,128\,231\,1$^d$&-0.588\,991\,112\,091\,6$^d$&-0.580\,721\,828\,1 &-0.572\,877\,277\,2\\ 
 2&-0.597\,299\,643\,396\,5&-0.588\,610\,829\,493&-0.580\,359\,195\,31&-0.572\,531\,809\,9\\
  &-0.597\,299\,643\,396\,469$^e$&&&\\
  &-0.597\,299\,643\,4     &-0.588\,610\,829\,6  &-0.580\,359\,195\,3 &-0.572\,531\,810\,4\\ 
 3&-0.596\,704\,882\,815\,2&-0.588\,043\,264\,274&-0.579\,818\,002\,15&-0.572\,016\,268\,8\\ 
  &-0.596\,704\,882\,815\,162$^e$&&&\\ 
  &-0.596\,704\,882\,6     &-0.588\,043\,264\,2  &-0.579\,818\,002\,0 &-0.572\,016\,269\,3\\ 
 4&-0.595\,917\,342\,277\,4&-0.587\,291\,784\,496&-0.579\,101\,495\,70&-0.571\,333\,785\,6\\
  &-0.595\,917\,342\,277\,360$^e$&&&\\ 
  &-0.595\,917\,342\,4     &-0.587\,291\,784\,7  &-0.579\,101\,495\,9 &-0.571\,333\,786\,3\\ 
 5&-0.594\,941\,578\,904\,4&-0.586\,360\,780\,057&-0.578\,213\,907\,45&-0.570\,488\,441\,9\\
  &-0.594\,941\,578\,904\,345$^e$&&&\\ 
  &-0.594\,941\,579\,1     &-0.586\,360\,780\,0  &-0.578\,213\,907\,7 &-0.570\,488\,442\,6\\ 
 6&-0.593\,783\,127\,985\,9&-0.585\,255\,582\,110&-0.577\,160\,375\,39&-0.569\,485\,193\,1\\
  &-0.593\,783\,127\,985\,818$^e$&&&\\
  &-0.593\,783\,128\,1     &-0.585\,255\,582\,1  &-0.577\,160\,375\,4 &-0.569\,485\,193\,6\\ 
 7&-0.592\,448\,405\,873\,9&-0.583\,982\,369\,086&-0.575\,946\,853\,12&-0.568\,329\,780\,6\\
  &-0.592\,448\,405\,873\,834$^e$&&&\\
  &-0.592\,448\,405\,7     &-0.583\,982\,368\,9  &-0.575\,946\,853\,0 &-0.568\,329\,781\,2\\ 
 8&-0.590\,944\,602\,712\,5&-0.582\,548\,063\,063&-0.574\,580\,009\,81&-0.567\,028\,635\,2\\
  &-0.590\,944\,602\,712\,443$^e$&&&\\ 
  &-0.590\,944\,602\,8     &-0.582\,548\,062\,8  &-0.574\,580\,009\,8 &-0.567\,028\,635\,9\\ 
 9&-0.589\,279\,568\,877\,4&-0.580\,960\,220\,249&-0.573\,067\,124\,61&-0.565\,588\,775\,5\\
  &-0.589\,279\,568\,877\,377$^e$&&&\\
  &-0.589\,279\,568\,9     &-0.580\,960\,220\,1  &-0.573\,067\,124\,7 &-0.565\,588\,776\,1\\ 
10&-0.587\,461\,698\,867\,7&-0.579\,226\,919\,228&-0.571\,415\,979\,14&-0.564\,017\,704\,8\\
  &-0.587\,461\,698\,867\,668$^e$&&&\\ 
  &-0.587\,461\,698\,6     &-0.579\,226\,919\,1  &-0.571\,415\,978\,9 &-0.564\,017\,705\,2\\ 
11&-0.585\,499\,816\,071\,2&-0.577\,356\,650\,307&-0.569\,634\,751\,33&-0.562\,323\,308\,5\\
  &-0.585\,499\,816\,071\,126$^e$&&&\\ 
  &-0.585\,499\,815\,8     &-0.577\,356\,650\,1  &-0.569\,634\,751\,2 &-0.562\,323\,309\,4\\ 
12&-0.583\,403\,061\,368\,7&-0.575\,358\,208\,792&-0.567\,731\,913\,16&-0.560\,513\,757\,1\\
  &-0.583\,403\,061\,368\,632$^e$&&&\\
  &-0.583\,403\,061\,4     &-0.575\,358\,208\,6  &-0.567\,731\,913\,1 &-0.560\,513\,757\,6\\ 
13&-0.581\,180\,788\,002\,2&-0.573\,240\,594\,524&-0.565\,716\,134\,74&-0.558\,597\,414\,2\\ 
  &-0.581\,180\,788\,2     &-0.573\,240\,594\,4  &-0.565\,716\,134\,7 &-0.558\,597\,414\,9\\ 
14&-0.578\,842\,464\,558\,5&-0.571\,012\,919\,429&-0.563\,596\,196\,28&-0.556\,582\,752\,9\\ 
  &-0.578\,842\,464\,6     &-0.571\,012\,919\,6  &-0.563\,596\,196\,3 &-0.556\,582\,753\,8\\ 
15&-0.576\,397\,587\,359\,8&-0.568\,684\,324\,282&-0.561\,380\,909\,07&-0.554\,478\,280\,9\\ 
  &-0.576\,397\,587\,5     &-0.568\,684\,324\,3  &-0.561\,380\,909\,3 &-0.554\,478\,281\,5\\ 
16&-0.573\,855\,603\,031\,8&-0.566\,263\,905\,388&-0.559\,079\,046\,12&-0.552\,292\,474\,9\\ 
  &-0.573\,855\,603\,0     &-0.566\,263\,905\,6  &-0.559\,079\,046\,0 &-0.552\,292\,475\,8\\ 
17&-0.571\,225\,841\,568\,3&-0.563\,760\,651\,451&-0.556\,699\,282\,76&-0.550\,033\,725\,1\\ 
  &-0.571\,225\,841\,7     &-0.563\,760\,651\,3  &-0.556\,699\,282\,6 &-0.550\,033\,725\,9\\ 
18&-0.568\,517\,459\,835\,2&-0.561\,183\,390\,542&-0.554\,250\,147\,03&-0.547\,710\,288\,8\\ 
  &-0.568\,517\,459\,9     &-0.561\,183\,390\,4  &-0.554\,250\,147\,1 &-0.547\,710\,289\,7\\ 
19&-0.565\,739\,395\,161\,4&-0.558\,540\,746\,805&-0.551\,739\,979\,43&-0.545\,330\,253\,6\\ 
  &-0.565\,739\,395\,3     &-0.558\,540\,746\,9  &-0.551\,739\,979\,5 &-0.545\,330\,254\,2\\ 
20&-0.562\,900\,328\,449\,7&-0.555\,841\,106\,348&-0.549\,176\,901\,70&-0.542\,901\,508\,8\\ 
  &-0.562\,900\,328\,2     &-0.555\,841\,106\,4  &-0.549\,176\,901\,5 &-0.542\,901\,509\,3\\ 
21&-0.560\,008\,656\,092   &-0.553\,092\,591\,62 &-0.546\,568\,793\,72&-0.540\,431\,725\,4\\ 
  &-0.560\,008\,656\,0     &-0.553\,092\,591\,5  &-0.546\,568\,793\,7 &-0.540\,431\,725\,9\\ 
22&-0.557\,072\,469\,897   &-0.550\,303\,043\,53 &-0.543\,923\,278\,04&-0.537\,928\,343\,1\\ 
  &-0.557\,072\,469\,7     &-0.550\,303\,043\,5  &-0.543\,923\,277\,9 &-0.537\,928\,343\,6\\ 
23&-0.554\,099\,544\,194   &-0.547\,480\,010\,55 &-0.541\,247\,711\,25&-0.535\,398\,564\,6\\ 
  &-0.554\,099\,544\,3     &-0.547\,480\,010\,7  &-0.541\,247\,711\,3 &-0.535\,398\,565\,1\\ 
24&-0.551\,097\,329\,307   &-0.544\,630\,743\,97 &-0.538\,549\,181\,51&-0.532\,849\,356\,0\\ 
  &-0.551\,097\,329\,4     &-0.544\,630\,743\,8  &-0.538\,549\,181\,6 &-0.532\,849\,356\,6\\ 
25&-0.548\,072\,950\,602   &-0.541\,762\,198\,76 &-0.535\,834\,511\,66&-0.530\,287\,453\,0\\ 
  &-0.548\,072\,950\,7     &-0.541\,762\,198\,5  &-0.535\,834\,511\,6 &-0.530\,287\,453\,6\\ 
26&-0.545\,033\,212\,416   &-0.538\,881\,039\,26 &-0.533\,110\,267\,45&-0.527\,719\,373\,0\\ 
  &-0.545\,033\,212\,6     &-0.538\,881\,039\,1  &-0.533\,110\,267\,4 &-0.527\,719\,373\,7\\ 
27&-0.541\,984\,606\,220   &-0.535\,993\,649\,26 &-0.530\,382\,770\,54&-0.525\,151\,432\,3\\ 
  &-0.541\,984\,606\,0     &-0.535\,993\,649\,4  &-0.530\,382\,770\,4 &-0.525\,151\,432\,7\\ 
28&-0.538\,933\,322\,483   &-0.533\,106\,146\,10 &-0.527\,658\,116\,05&-0.522\,589\,769\,2\\ 
  &-0.538\,933\,322\,6     &-0.533\,106\,146\,2  &-0.527\,658\,116\,0 &-0.522\,589\,769\,4\\ 
29&-0.535\,885\,265\,823   &-0.530\,224\,398\,43 &-0.524\,942\,194\,62&-0.520\,040\,373\,4\\ 
  &-0.535\,885\,265\,5     &-0.530\,224\,398\,5  &-0.524\,942\,194\,4 &-0.520\,040\,373\,6\\ 
30&-0.532\,846\,073\,113   &-0.527\,354\,047\,63 &-0.522\,240\,719\,44&-0.517\,509\,122\,6\\ 
  &-0.532\,846\,072\,8     &-0.527\,354\,047\,6  &-0.522\,240\,719\,5 &-0.517\,509\,123\,0\\ 
31&-0.529\,821\,134\,389   &-0.524\,500\,532\,93 &-0.519\,559\,258\,59&-0.515\,001\,827\,2\\ 
  &-0.529\,821\,134\,1     &-0.524\,500\,532\,9  &-0.519\,559\,258\,3 &-0.515\,001\,827\,3\\ 
32&-0.526\,815\,616\,499   &-0.521\,669\,120\,58 &-0.516\,903\,274\,04&-0.512\,524\,287\,4\\ 
  &-0.526\,815\,616\,4     &-0.521\,669\,120\,4  &-0.516\,903\,273\,9 &-0.512\,524\,287\,6\\ 
33&-0.523\,834\,489\,683   &-0.518\,864\,937\,83 &-0.514\,278\,168\,95&-0.510\,082\,365\,5\\ 
  &-0.523\,834\,489\,6     &-0.518\,864\,937\,6  &-0.514\,278\,168\,7 &-0.510\,082\,365\,8\\ 
34&-0.520\,882\,557\,405   &-0.516\,093\,012\,69 &-0.511\,689\,346\,16&-0.507\,682\,081\,6\\ 
  &-0.520\,882\,557\,5     &-0.516\,093\,012\,6  &-0.511\,689\,346\,0 &-0.507\,682\,081\,5\\ 
35&-0.517\,964\,490\,152   &-0.513\,358\,321\,48 &-0.509\,142\,282\,33&-0.505\,329\,742\,9\\ 
  &-0.517\,964\,490\,0     &-0.513\,358\,321\,6  &-0.509\,142\,282\,2 &-0.505\,329\,743\,1\\ 
36&-0.515\,084\,864\,210   &-0.510\,665\,846\,79 &-0.506\,642\,625\,18&-0.503\,032\,128\,7\\ 
  &-0.515\,084\,864\,2     &-0.510\,665\,846\,9  &-0.506\,642\,625\,2 &-0.503\,032\,128\,6\\ 
37&-0.512\,248\,207\,112   &-0.508\,020\,650\,51 &-0.504\,196\,326\,01&-0.500\,796\,767\,3\\ 
  &-0.512\,248\,207\,0     &-0.508\,020\,650\,5  &-0.504\,196\,326\,0 &-0.500\,796\,767\,3\\ \cline{5-5}
38&-0.509\,459\,052\,264   &-0.505\,427\,968\,98 &-0.501\,809\,829\,30&-0.498\,632\,382\,9\\ 
  &-0.509\,459\,052\,3     &-0.505\,427\,969\,0  &-0.501\,809\,829\,2 &-0.498\,632\,382\,7\\ \cline{4-4}
39&-0.506\,722\,006\,729   &-0.502\,893\,342\,59 &-0.499\,490\,360\,08&-0.496\,549\,678\,8\\ 
  &-0.506\,722\,006\,5     &-0.502\,893\,342\,6  &-0.499\,490\,360\,2 &-0.496\,549\,678\,7\\ 
40&-0.504\,041\,838\,520   &-0.500\,422\,801\,25 &-0.497\,246\,392\,46&-0.494\,562\,907\,4\\ 
  &-0.504\,041\,838\,5     &-0.500\,422\,801\,0  &-0.497\,246\,392\,5 &-0.494\,562\,907\,4\\ \cline{3-3}
41&-0.501\,423\,593\,974   &-0.498\,023\,146\,52 &-0.495\,088\,491\,10&-0.492\,693\,76    \\ 
  &-0.501\,423\,593\,7     &-0.498\,023\,146\,5  &-0.495\,088\,491\,1 &-0.492\,693\,77    \\ \cline{2-2} 
42&-0.498\,872\,763\,78    &-0.495\,702\,414\,99 &-0.493\,031\,052\,7 &-0.490\,987        \\ 
  &-0.498\,872\,763\,6     &-0.495\,702\,414\,9  &-0.493\,031\,052\,9 &-0.490\,99         \\ 
43&-0.496\,395\,532\,49    &-0.493\,470\,721\,23 &-0.491\,096\,94     &\\ 
  &-0.496\,395\,532\,4     &-0.493\,470\,721\,0  &-0.491\,097\,0      &\\ 
44&-0.493\,999\,183\,31    &-0.491\,342\,046\,6  &&\\
  &-0.493\,999\,183\,4     &-0.491\,342\,046\,7  &&\\
45&-0.491\,692\,825\,67    &-0.489\,339\,4       &&\\
  &-0.491\,692\,825\,8     &-0.489\,339          &&\\
46&-0.489\,488\,926\,5     &&&\\
  &-0.489\,488\,927        &&&\\
47&-0.487\,407\,9          &&&\\
  &-0.487\,408             &&&\\
\end{longtable}
\end{center}             
The energy for the ground state $(L^\pi,v) = (0^+,0)$ has been improved in a series of papers~
\cite{Mo99,HBG00,F02,YZL03}. 
The best known values were determined with about 18 digits by Yan \etal \cite{YZ04} 
and about 25 digits by Hijikata \etal \cite{HNN09}, 
with different values for the masses of the proton and deuteron. 
Our accuracy is about $10^{-13}$. 
For the $(0^+,1)$, $(0^+,2)$ and $(0^+,3)$ vibrational excited states, 
the accuracies are about $10^{-11}$, $10^{-10}$ and $10^{-9}$, respectively. 
 
\begin{figure}[hbt]
\setlength{\unitlength}{1mm}
\begin{picture}(140,70) (-20,10) 
\put(0,0){\mbox{\scalebox{1.5}{\includegraphics{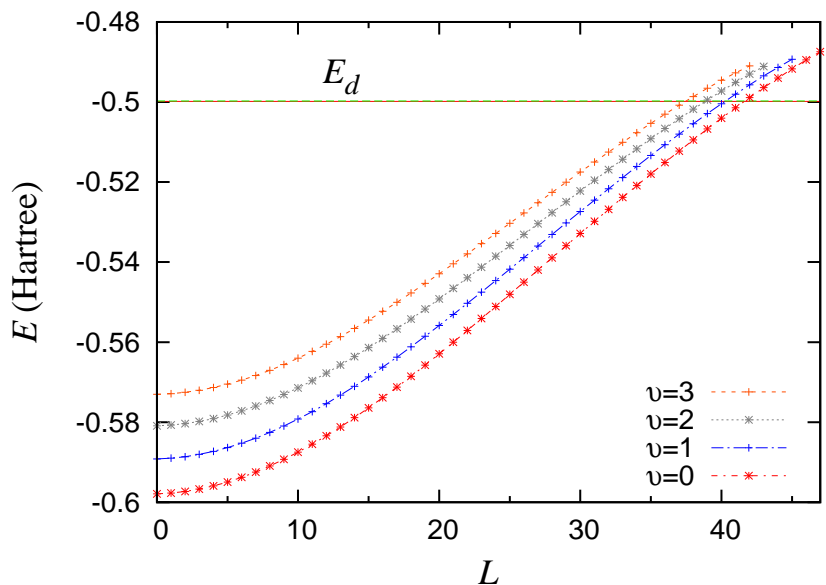}}}}
\end{picture} \\
\caption{Four lowest $\Sigma_g$ rotational bands of the HD$^+$ molecular ion.
The two dissociation energies $E_d^{{\rm\, H+D}^+}$ and $E_d^{{\rm\, D+H}^+}$ 
can not be distinguished.}
\label{fig:1}
\end{figure}

The energy of the lowest $L = 1$ level is known with about 18 digits \cite{YZL03}. 
Results with the same accuracy (close to 18 digits) are available for $L = 2-12$ and $v = 0$ \cite{YZ04}. 
Our error for the lowest vibrational energy remains smaller than 
$10^{-13}$ for all these states.
When comparing the rest of our results with those of Moss \cite{Mo93HD}, 
one observes that both works agree very well. 
The present energies are a little more accurate for $v=0$ and a little less accurate for $v=2$ and $3$. 
But, in addition, the Lagrange-mesh method provides easy-to-use accurate wave functions. 
Fig.~\ref{fig:1} shows the obtained spectrum.

Energies of four vibrational levels $v=0-3$ and six rotational levels $L=0-5$ 
calculated with the CODATA 2006 values $m_p=1836.152\,672\,47$ and $m_d=3670.482\,965\,4$ 
are presented in Table~\ref{tab:2}. 
Each first line displays Lagrange-mesh results obtained using the set of parameters 
$N_x=N_y=40$, $N_z=20$ and $h_x=h_y=0.08$, $h_z=0.5$. 
Each second line contains rounded results of Tian \etal \cite{TTZY12}. 
Their ground-state energy agrees with the more accurate result of \cite{HNN09}. 
When comparing the results obtained with two different mass conventions 
in Tables \ref{tab:1} and \ref{tab:2}, 
one observes that the difference in the energies appears at the tenth digit.

\begin{table}[hbt]
\centering{
\caption{Comparison of energies for six rotational levels $L=0-5$ 
and four vibrational levels $v=0-3$. 
The mass conventions for the proton and deuteron are 
$m_p=1836.15267247$ and $m_d=3670.4829654$, respectively. 
The Lagrange-mesh results obtained with $N_x = N_y = 40$, $N_z = 20$ and 
$h_x = h_y = 0.08$, $h_z = 0.5$ are presented in each first line. 
Each second line displays rounded results of Tian \etal \cite{TTZY12}.} 
\label{tab:2} 
\resizebox{16cm}{!}{
\begin{tabular}{lllll}
\hline
\hline
$L$& $v=0$          & $v=1$           &  $v=2$          &$v=3$\\
\hline
 0&-0.597\,897\,968\,609\,0 &-0.589\,181\,829\,557   &-0.580\,903\,700\,22 &-0.573\,050\,546\,5 \\ 
  &-0.597\,897\,968\,608\,95&-0.589\,181\,829\,556\,7&-0.580\,903\,700\,218&-0.573\,050\,546\,55\\ 
 1&-0.597\,698\,128\,192\,1 &-0.588\,991\,111\,992   &-0.580\,721\,828\,12 &-0.572\,877\,277\,1 \\ 
  &-0.597\,698\,128\,192\,13&-0.588\,991\,111\,991\,8&-0.580\,721\,828\,121&-0.572\,877\,277\,09\\ 
 2&-0.597\,299\,643\,351\,7 &-0.588\,610\,829\,390   &-0.580\,359\,195\,20 &-0.572\,531\,810\,3 \\ 
  &-0.597\,299\,643\,351\,68&-0.588\,610\,829\,389\,6&-0.580\,359\,195\,200&-0.572\,531\,810\,33\\ 
 3&-0.596\,704\,882\,761\,8 &-0.588\,043\,264\,163   &-0.579\,818\,002\,03 &-0.572\,016\,269\,2 \\ 
  &-0.596\,704\,882\,761\,78&-0.588\,043\,264\,162\,6&-0.579\,818\,002\,028&-0.572\,016\,269\,23\\ 
 4&-0.595\,917\,342\,212\,7 &-0.587\,291\,784\,374   &-0.579\,101\,495\,56 &-0.571\,333\,786\,0 \\ 
  &-0.595\,917\,342\,212\,67&-0.587\,291\,784\,373\,9&-0.579\,101\,495\,565&-0.571\,333\,786\,05\\ 
 5&-0.594\,941\,578\,825\,8 &-0.586\,360\,779\,923   &-0.578\,213\,907\,30 &-0.570\,488\,442\,3 \\ 
  &-0.594\,941\,578\,825\,77&-0.586\,360\,779\,922\,8&-0.578\,213\,907\,303&-0.570\,488\,442\,33\\ 
\hline
\end{tabular}
}}
\end{table}
\begin{table}[hbt]
\centering{
\caption{Convergence of the energies and transition probabilities 
as a function of the numbers $N_x = N_y$ and $N_z$ of mesh points. 
Two cases are shown: $(2^+,1) \rightarrow (3^-,0)$ where $L_f = L_i + 1$ (upper set) 
and $(40^+,0) \rightarrow (39^-,0)$ where $L_f = L_i - 1$ (lower set). 
The scale factors are $h_x = h_y = 0.11$ and $h_z = 0.4$.} 
\label{tab:3} 
\resizebox{16cm}{!}{
\begin{tabular}{clllll}
\hline
$N_{x,y}$&$N_z$&$E_i (2^+,1)$&$E_f (3^-,0)$&$W_0$ ($10$\,s$^{-1}$)&$W$ ($10$\,s$^{-1}$)\\
\hline
38&14&-0.588\,610\,829\,483\,454&-0.596\,704\,882\,814\,981&1.119\,849\,260&1.119\,957\,466\\
38&20&-0.588\,610\,829\,483\,336&-0.596\,704\,882\,814\,924&1.119\,849\,261&1.119\,957\,466\\
40&14&-0.588\,610\,829\,492\,821&-0.596\,704\,882\,815\,199&1.119\,849\,257&1.119\,957\,462\\
40&20&-0.588\,610\,829\,492\,703&-0.596\,704\,882\,815\,146&1.119\,849\,257&1.119\,957\,462\\
42&14&-0.588\,610\,829\,494\,525&-0.596\,704\,882\,815\,244&1.119\,849\,256&1.119\,957\,461\\
44&14&-0.588\,610\,829\,494\,882&-0.596\,704\,882\,815\,238&1.119\,849\,254&1.119\,957\,459\\
46&14&-0.588\,610\,829\,494\,935&-0.596\,704\,882\,815\,237&1.119\,849\,260&1.119\,957\,465\\
\hline
$N_{x,y}$&$N_z$&$E_i (40^+,0)$&$E_f (39^-,0)$&$W_0$ ($10$\,s$^{-1}$)&$W$ ($10$\,s$^{-1}$)\\
\hline
38& 14& -0.504\,041\,838\,520\,152&-0.506\,722\,006\,729\,020&6.993\,933\,079&6.992\,287\,324\\ 
38& 20& -0.504\,041\,838\,516\,907&-0.506\,722\,006\,726\,162&6.993\,933\,080&6.992\,287\,325\\ 
40& 14& -0.504\,041\,838\,520\,210&-0.506\,722\,006\,729\,109&6.993\,933\,079&6.992\,287\,324\\ 
40& 20& -0.504\,041\,838\,516\,956&-0.506\,722\,006\,726\,242&6.993\,933\,081&6.992\,287\,326\\ 
42& 14& -0.504\,041\,838\,520\,210&-0.506\,722\,006\,729\,114&6.993\,933\,079&6.992\,287\,324\\ 
44& 14& -0.504\,041\,838\,520\,210&-0.506\,722\,006\,729\,107&6.993\,933\,079&6.992\,287\,324\\ 
46& 14& -0.504\,041\,838\,520\,207&-0.506\,722\,006\,729\,118&6.993\,933\,079&6.992\,287\,324\\ 
\hline
\end{tabular}
}}
\end{table}
Table \ref{tab:3} present convergence tests as a function of the number of mesh points for two 
different sets of initial and final levels. 
The scaling parameters are $h_x = h_y = 0.11$ and $h_z = 0.4$. 
The initial and final energies as well as transition probabilities are displayed. 
The transition probability $W_0$ is obtained by restricting \eref{4.8} to $\kappa = 0$ 
while $W_1$ corresponds to $\kappa \le 1$. 

In general, the $\kappa = 1$ contributions have an importance smaller than 0.02~\%. 
With respect to $N_z$, a 12-digit convergence of the energies and 
a 10-digit convergence of the probabilities is already obtained for $N_z = 14$. 
The convergence with respect to $N$ is slower. 
Since the convergence is exponential, one can estimate that the relative accuracy 
on $W$ with $N_x = N_y = 40$ and $N_z = 14$ is about $10^{-8}$ for the two cases 
$(2^+,1)\rightarrow(3^-,0)$ and $(40^+,0)\rightarrow(39^-,0)$. 
Further similar tests have been performed for other transitions. 

The convergence of the transition probabilities with respect to $K_{\rm max}$ 
can be studied by comparing the results for $K_{\rm max} = 2$ 
with results from wave functions truncated at $K_{\rm max} = 0$ and $K_{\rm max}= 1$. 
The relative error when $K_{\rm max} = 0$ is smaller than 9 \% for all considered transitions  
while the error for $K_{\rm max} = 1$ is smaller than $5\times10^{-6}$. 
By extrapolation, we estimate that the relative error on the present transition probabilities 
obtained with $K_{\rm max}= 2$ should be smaller than $10^{-7}$. 

Table \ref{tab:4} presents transition probabilities per second within a same rotational band, 
$L_f = L_i - 1$ and $v_f = v_i \le 3$. 
Some transition probabilities involving quasibound levels are also included. 
We limit the number of significant figures to six. 
The probabilities increase slowly with $L$ with a maximum at $L_i = 36$, 35, 33 and 32 
for $v_i = 0$, 1, 2 and 3, respectively. 
This is due to a maximum of the energy differences around $L_i = 27$. 
The maximum of the transition probabilities is shifted toward higher $L_i$ values 
by a steady increase of the reduced matrix elements. 

\begin{center}
\begin{longtable}{rllll}
\caption{Dipole transition probabilities per second $W$ for transitions 
between levels of a same rotational band ($v_f = v_i$, $L_f = L_i - 1$). 
Results are given with five digits followed by the power of 10.} 
\label{tab:4} \\
\\[-4.9ex]
\hline
$L_i$&$v_i=0$&$v_i=1$&$v_i=2$&$v_i=3$\\
\hline
\endfirsthead
\multicolumn{4}{c}{{\tablename} \thetable{} -- Continuation}\\
\hline
$L_i$&$v_i=0$&$v_i=1$&$v_i=2$&$v_i=3$\\
\hline
\endhead
\hline
\multicolumn{4}{l}{{Continued on Next Page\ldots}}\\
\endfoot
\hline
\endlastfoot
 1& 6.697\,53\m3& 6.501\,48\m3& 6.282\,42\m3& 6.042\,37\m3\\ 
 2& 6.390\,50\m2& 6.201\,76\m2& 5.991\,21\m2& 5.760\,79\m2\\ 
 3& 2.287\,55\m1& 2.219\,00\m1& 2.142\,73\m1& 2.059\,43\m1\\ 
 4& 5.544\,41\m1& 5.374\,96\m1& 5.187\,08\m1& 4.982\,44\m1\\ 
 5& 1.087\,75\p0& 1.053\,69\p0& 1.016\,08\p0& 9.752\,57\m1\\ 
 6& 1.867\,43\p0& 1.807\,27\p0& 1.741\,18\p0& 1.669\,70\p0\\ 
 7& 2.923\,00\p0& 2.825\,80\p0& 2.719\,60\p0& 2.605\,20\p0\\ 
 8& 4.274\,11\p0& 4.127\,03\p0& 3.967\,21\p0& 3.795\,82\p0\\ 
 9& 5.930\,37\p0& 5.718\,75\p0& 5.490\,12\p0& 5.246\,04\p0\\ 
10& 7.891\,61\p0& 7.599\,17\p0& 7.285\,04\p0& 6.951\,22\p0\\ 
11& 1.014\,86\p1& 9.757\,66\p0& 9.340\,13\p0& 8.898\,44\p0\\ 
12& 1.268\,38\p1& 1.217\,57\p1& 1.163\,60\p1& 1.106\,75\p1\\ 
13& 1.547\,29\p1& 1.482\,81\p1& 1.414\,69\p1& 1.343\,24\p1\\ 
14& 1.848\,58\p1& 1.768\,45\p1& 1.684\,22\p1& 1.596\,24\p1\\ 
15& 2.168\,79\p1& 2.071\,04\p1& 1.968\,75\p1& 1.862\,34\p1\\ 
16& 2.504\,17\p1& 2.386\,84\p1& 2.264\,61\p1& 2.137\,91\p1\\ 
17& 2.850\,78\p1& 2.712\,00\p1& 2.568\,01\p1& 2.419\,26\p1\\ 
18& 3.204\,61\p1& 3.042\,59\p1& 2.875\,12\p1& 2.702\,64\p1\\ 
19& 3.561\,66\p1& 3.374\,73\p1& 3.182\,18\p1& 2.984\,43\p1\\ 
20& 3.918\,04\p1& 3.704\,67\p1& 3.485\,56\p1& 3.261\,11\p1\\ 
21& 4.270\,03\p1& 4.028\,82\p1& 3.781\,82\p1& 3.529\,37\p1\\ 
22& 4.614\,11\p1& 4.343\,83\p1& 4.067\,74\p1& 3.786\,10\p1\\ 
23& 4.947\,04\p1& 4.646\,58\p1& 4.340\,33\p1& 4.028\,46\p1\\ 
24& 5.265\,82\p1& 4.934\,24\p1& 4.596\,88\p1& 4.253\,82\p1\\ 
25& 5.567\,77\p1& 5.204\,24\p1& 4.834\,94\p1& 4.459\,82\p1\\ 
26& 5.850\,49\p1& 5.454\,29\p1& 5.052\,30\p1& 4.644\,32\p1\\ 
27& 6.111\,85\p1& 5.682\,36\p1& 5.247\,00\p1& 4.805\,39\p1\\ 
28& 6.349\,99\p1& 5.886\,67\p1& 5.417\,31\p1& 4.941\,29\p1\\ 
29& 6.563\,34\p1& 6.065\,68\p1& 5.561\,68\p1& 5.050\,45\p1\\ 
30& 6.750\,52\p1& 6.218\,03\p1& 5.678\,73\p1& 5.131\,40\p1\\ 
31& 6.910\,39\p1& 6.342\,55\p1& 5.767\,20\p1& 5.182\,72\p1\\ 
32& 7.041\,98\p1& 6.438\,21\p1& 5.825\,93\p1& 5.203\,03\p1\\ 
33& 7.144\,46\p1& 6.504\,06\p1& 5.853\,79\p1& 5.190\,83\p1\\ 
34& 7.217\,13\p1& 6.539\,23\p1& 5.849\,58\p1& 5.144\,45\p1\\ 
35& 7.259\,38\p1& 6.542\,84\p1& 5.812\,01\p1& 5.061\,89\p1\\ 
36& 7.270\,62\p1& 6.513\,96\p1& 5.739\,55\p1& 4.940\,53\p1\\ 
37& 7.250\,27\p1& 6.451\,50\p1& 5.630\,25\p1& 4.776\,80\p1\\ 
38& 7.197\,68\p1& 6.354\,10\p1& 5.481\,46\p1& 4.565\,42\p1\\ 
39& 7.112\,03\p1& 6.219\,96\p1& 5.289\,43\p1& 4.297\,96\p1\\ 
40& 6.992\,29\p1& 6.046\,57\p1& 5.048\,41\p1& 3.959\,40\p1\\ 
41& 6.836\,99\p1& 5.830\,20\p1& 4.748\,90\p1& 3.516\,7\s\p1\\ 
42& 6.644\,04\p1& 5.565\,01\p1& 4.373\,14\p1& 2.8\sm\s\s\p1\\ 
43& 6.410\,27\p1& 5.241\,07\p1& 3.878\sm\p1& \\
44& 6.130\,62\p1& 4.839\,0\s \p1& &\\
45& 5.796\,43\p1& 4.307\sm\p1& &\\
46& 5.390\,55\p1& &&\\
47& 4.86\sm\s\p1& &&\\
\end{longtable}
\end{center}

The probabilities per second for other transitions are displayed in Table \ref{tab:5}. 
The columns correspond to transitions between different vibrational levels ($v_i\rightarrow v_f$). 
For each $L_i>0$, the first line corresponds to $L_f = L_i - 1$ 
and the second line to $L_f = L_i+1$.

The strongest transition from each level occurs in general towards the nearest vibrational level 
($v_f = v_i - 1$). 
For $L_f = L_i - 1$ exceptions can be found between $L_i = 23$ and $L_i = 28$. 

\begin{center}
\begin{longtable}{lllllll}
\caption{Dipole transition probabilities per second $W$ for transitions 
between different vibrational quantum numbers $(v_i \ne v_f)$. 
For $L_i \ge 1$, each first and second lines correspond to 
$L_f = L_i - 1$ and $L_f = L_i + 1$, respectively.} 
\label{tab:5}\\
\\[-4.9ex]
\hline
$L_i$&$(1\rightarrow\,0)$&$(2\rightarrow\,0)$&$(2\rightarrow\,1)$
     &$(3\rightarrow\,0)$&$(3\rightarrow\,1)$&$(3\rightarrow\,2)$\\
\hline
\endfirsthead
\multicolumn{7}{c}{{\tablename} \thetable{} -- Continuation}\\
\hline
$L_i$&$(1\rightarrow\,0)$&$(2\rightarrow\,0)$&$(2\rightarrow\,1)$
     &$(3\rightarrow\,0)$&$(3\rightarrow\,1)$&$(3\rightarrow\,2)$\\
\hline
\endhead
\hline
\multicolumn{7}{l}{{Continued on Next Page\ldots}}\\
\endfoot
\hline
\endlastfoot
 0& 1.831\,21\p1& 2.018\,40\p0& 3.208\,68\p1& 3.023\,44\m1& 5.190\,59\p0& 4.206\,38\p1\\ 
 1& 5.811\,13\p0& 6.888\,29\m1& 1.013\,57\p1& 1.071\,82\m1& 1.763\,01\p0& 1.322\,54\p1\\ 
  & 1.237\,03\p1& 1.315\,61\p0& 2.171\,76\p1& 1.933\,67\m1& 3.390\,82\p0& 2.852\,55\p1\\ 
 2& 6.723\,81\p0& 8.270\,79\m1& 1.169\,62\p1& 1.311\,63\m1& 2.111\,51\p0& 1.521\,98\p1\\ 
  & 1.119\,96\p1& 1.149\,78\p0& 1.969\,60\p1& 1.658\,29\m1& 2.969\,72\p0& 2.591\,40\p1\\ 
 3& 6.889\,60\p0& 8.799\,00\m1& 1.194\,93\p1& 1.422\,15\m1& 2.240\,45\p0& 1.550\,19\p1\\ 
  & 1.065\,50\p1& 1.056\,41\p0& 1.876\,63\p1& 1.495\,27\m1& 2.734\,11\p0& 2.472\,68\p1\\ 
 4& 6.775\,74\p0& 8.990\,27\m1& 1.171\,39\p1& 1.480\,90\m1& 2.282\,88\p0& 1.514\,54\p1\\ 
  & 1.027\,89\p1& 9.847\,15\m1& 1.812\,71\p1& 1.368\,10\m1& 2.553\,50\p0& 2.391\,37\p1\\ 
 5& 6.515\,50\p0& 8.988\,10\m1& 1.122\,42\p1& 1.508\,92\m1& 2.275\,82\p0& 1.445\,85\p1\\ 
  & 9.949\,76\p0& 9.215\,37\m1& 1.756\,56\p1& 1.257\,01\m1& 2.394\,08\p0& 2.319\,61\p1\\ 
 6& 6.165\,67\p0& 8.851\,31\m1& 1.058\,05\p1& 1.514\,51\m1& 2.234\,53\p0& 1.357\,35\p1\\ 
  & 9.624\,59\p0& 8.623\,82\m1& 1.700\,65\p1& 1.155\,23\m1& 2.244\,34\p0& 2.247\,52\p1\\ 
 7& 5.757\,71\p0& 8.611\,62\m1& 9.838\,57\p0& 1.501\,97\m1& 2.167\,28\p0& 1.256\,48\p1\\ 
  & 9.286\,79\p0& 8.056\,00\m1& 1.642\,07\p1& 1.060\,18\m1& 2.100\,06\p0& 2.171\,34\p1\\ 
 8& 5.312\,63\p0& 8.290\,32\m1& 9.035\,84\p0& 1.474\,11\m1& 2.079\,67\p0& 1.148\,21\p1\\ 
  & 8.930\,86\p0& 7.506\,56\m1& 1.579\,91\p1& 9.708\,90\m2& 1.959\,92\p0& 2.089\,87\p1\\ 
 9& 4.846\,21\p0& 7.904\,17\m1& 8.200\,39\p0& 1.433\,22\m1& 1.976\,07\p0& 1.036\,28\p1\\ 
  & 8.556\,52\p0& 6.974\,91\m1& 1.514\,14\p1& 8.870\,33\m2& 1.823\,83\p0& 2.003\,15\p1\\ 
10& 4.371\,14\p0& 7.467\,63\m1& 7.354\,66\p0& 1.381\,32\m1& 1.860\,26\p0& 9.236\,71\p0\\ 
  & 8.166\,08\p0& 6.462\,38\m1& 1.445\,20\p1& 8.085\,26\m2& 1.692\,20\p0& 1.911\,81\p1\\ 
11& 3.897\,90\p0& 6.993\,74\m1& 6.517\,14\p0& 1.320\,36\m1& 1.735\,62\p0& 8.128\,19\p0\\ 
  & 7.763\,10\p0& 5.970\,90\m1& 1.373\,76\p1& 7.353\,51\m2& 1.565\,57\p0& 1.816\,80\p1\\ 
12& 3.435\,21\p0& 6.494\,43\m1& 5.703\,04\p0& 1.252\,20\m1& 1.605\,22\p0& 7.057\,14\p0\\ 
  & 7.351\,66\p0& 5.502\,38\m1& 1.300\,59\p1& 6.674\,79\m2& 1.444\,52\p0& 1.719\,16\p1\\ 
13& 2.990\,23\p0& 5.980\,53\m1& 4.924\,74\p0& 1.178\,64\m1& 1.471\,82\p0& 6.039\,54\p0\\ 
  & 6.935\,91\p0& 5.058\,37\m1& 1.226\,45\p1& 6.048\,29\m2& 1.329\,49\p0& 1.619\,98\p1\\ 
14& 2.568\,73\p0& 5.461\,75\m1& 4.192\,03\p0& 1.101\,35\m1& 1.337\,89\p0& 5.087\,89\p0\\ 
  & 6.519\,78\p0& 4.639\,95\m1& 1.152\,08\p1& 5.472\,64\m2& 1.220\,83\p0& 1.520\,26\p1\\ 
15& 2.175\,15\p0& 4.946\,64\m1& 3.512\,37\p0& 1.021\,87\m1& 1.205\,57\p0& 4.211\,48\p0\\ 
  & 6.106\,87\p0& 4.247\,70\m1& 1.078\,14\p1& 4.945\,96\m2& 1.118\,72\p0& 1.420\,96\p1\\ 
16& 1.812\,75\p0& 4.442\,50\m1& 2.891\,08\p0& 9.416\,16\m2& 1.076\,68\p0& 3.416\,77\p0\\ 
  & 5.700\,36\p0& 3.881\,73\m1& 1.005\,23\p1& 4.465\,93\m2& 1.023\,24\p0& 1.322\,90\p1\\ 
17& 1.483\,74\p0& 3.955\,44\m1& 2.331\,58\p0& 8.617\,93\m2& 9.527\,21\m1& 2.707\,67\p0\\ 
  & 5.302\,93\p0& 3.541\,75\m1& 9.338\,52\p0& 4.029\,91\m2& 9.343\,56\m1& 1.226\,78\p1\\ 
18& 1.189\,37\p0& 3.490\,38\m1& 1.835\,63\p0& 7.834\,48\m2& 8.348\,91\m1& 2.085\,92\p0\\ 
  & 4.916\,82\p0& 3.227\,10\m1& 8.644\,30\p0& 3.635\,06\m2& 8.519\,31\m1& 1.133\,21\p1\\ 
19& 9.300\,83\m1& 3.051\,13\m1& 1.403\,61\p0& 7.074\,42\m2& 7.241\,06\m1& 1.551\,43\p0\\ 
  & 4.543\,81\p0& 2.936\,89\m1& 7.973\,01\p0& 3.278\,43\m2& 7.757\,60\m1& 1.042\,65\p1\\ 
20& 7.056\,51\m1& 2.640\,47\m1& 1.034\,70\p0& 6.344\,67\m2& 6.210\,23\m1& 1.102\,59\p0\\ 
  & 4.185\,27\p0& 2.670\,00\m1& 7.327\,27\p0& 2.957\,05\m2& 7.055\,78\m1& 9.554\,88\p0\\ 
21& 5.152\,83\m1& 2.260\,29\m1& 7.271\,47\m1& 5.650\,50\m2& 5.260\,72\m1& 7.365\,97\m1\\ 
  & 3.842\,21\p0& 2.425\,18\m1& 6.709\,03\p0& 2.668\,01\m2& 6.410\,81\m1& 8.720\,06\p0\\ 
22& 3.577\,53\m1& 1.911\,67\m1& 4.784\,72\m1& 4.995\,75\m2& 4.394\,88\m1& 4.497\,41\m1\\ 
  & 3.515\,30\p0& 2.201\,11\m1& 6.119\,64\p0& 2.408\,48\m2& 5.819\,42\m1& 7.924\,01\p0\\ 
23& 2.315\,08\m1& 1.595\,05\m1& 2.856\,59\m1& 4.382\,95\m2& 3.613\,40\m1& 2.376\,56\m1\\ 
  & 3.204\,93\p0& 1.996\,41\m1& 5.559\,92\p0& 2.175\,80\m2& 5.278\,15\m1& 7.168\,03\p0\\ 
24& 1.347\,63\m1& 1.310\,29\m1& 1.453\,25\m1& 3.813\,52\m2& 2.915\,68\m1& 9.554\,86\m2\\ 
  & 2.911\,27\p0& 1.809\,69\m1& 5.030\,28\p0& 1.967\,43\m2& 4.783\,51\m1& 6.452\,81\p0\\ 
25& 6.558\,42\m2& 1.056\,84\m1& 5.387\,07\m2& 3.287\,98\m2& 2.300\,09\m1& 1.839\,06\m2\\ 
  & 2.634\,28\p0& 1.639\,59\m1& 4.530\,75\p0& 1.781\,03\m2& 4.332\,03\m1& 5.778\,53\p0\\ 
26& 2.196\,64\m2& 8.338\,15\m2& 7.605\,85\m3& 2.806\,06\m2& 1.764\,28\m1& 1.083\,20\m3\\ 
  & 2.373\,77\p0& 1.484\,78\m1& 4.061\,08\p0& 1.614\,46\m2& 3.920\,33\m1& 5.145\,00\p0\\ 
27& 1.890\,54\m3& 6.401\,08\m2& 2.854\,72\m3& 2.366\,95\m2& 1.305\,39\m1& 3.859\,26\m2\\ 
  & 2.129\,43\p0& 1.344\,01\m1& 3.620\,79\p0& 1.465\,72\m2& 3.545\,13\m1& 4.551\,70\p0\\ 
28& 3.376\,70\m3& 4.744\,74\m2& 3.604\,48\m2& 1.969\,39\m2& 9.203\,45\m2& 1.260\,61\m1\\ 
  & 1.900\,86\p0& 1.216\,04\m1& 3.209\,25\p0& 1.333\,03\m2& 3.203\,29\m1& 3.997\,91\p0\\ 
29& 2.452\,65\m2& 3.356\,10\m2& 1.037\,79\m1& 1.611\,86\m2& 6.060\,22\m2& 2.588\,97\m1\\ 
  & 1.687\,58\p0& 1.099\,76\m1& 2.825\,68\p0& 1.214\,75\m2& 2.891\,86\m1& 3.482\,71\p0\\ 
30& 6.355\,78\m2& 2.222\,31\m2& 2.028\,93\m1& 1.292\,69\m2& 3.594\,97\m2& 4.328\,46\m1\\ 
  & 1.489\,07\p0& 9.940\,89\m2& 2.469\,24\p0& 1.109\,42\m2& 2.608\,03\m1& 3.005\,11\p0\\ 
31& 1.188\,34\m1& 1.331\,42\m2& 3.305\,09\m1& 1.010\,21\m2& 1.782\,39\m2& 6.440\,51\m1\\ 
  & 1.304\,80\p0& 8.980\,31\m2& 2.139\,05\p0& 1.015\,73\m2& 2.349\,16\m1& 2.564\,10\p0\\ 
32& 1.888\,89\m1& 6.730\,54\m3& 4.840\,74\m1& 7.628\,86\m3& 6.034\,10\m3& 8.890\,99\m1\\ 
  & 1.134\,22\p0& 8.106\,61\m2& 1.834\,21\p0& 9.325\,13\m3& 2.112\,78\m1& 2.158\,64\p0\\ 
33& 2.724\,50\m1& 2.391\,96\m3& 6.614\,02\m1& 5.494\,56\m3& 4.779\,8\s\m4& 1.165\,06\p0\\ 
  & 9.767\,86\m1& 7.311\,16\m2& 1.553\,88\p0& 8.587\,33\m3& 1.896\,55\m1& 1.787\,81\p0\\ 
34& 3.684\,60\m1& 2.506\,50\m4& 8.607\,15\m1& 3.691\,58\m3& 1.174\,86\m3& 1.469\,51\p0\\ 
  & 8.319\,91\m1& 6.585\,94\m2& 1.297\,25\p0& 7.934\,71\m3& 1.698\,26\m1& 1.450\,79\p0\\ 
35& 4.761\,07\m1& 3.023\,48\m4& 1.080\,69\p0& 2.219\,77\m3& 8.310\,15\m3& 1.800\,56\p0\\ 
  & 6.993\,53\m1& 5.923\,47\m2& 1.063\,62\p0& 7.359\,19\m3& 1.515\,80\m1& 1.146\,96\p0\\ 
36& 5.948\,53\m1& 2.601\,69\m3& 1.320\,50\p0& 1.090\,52\m3& 2.229\,92\m2& 2.156\,86\p0\\ 
  & 5.784\,41\m1& 5.316\,70\m2& 8.524\,10\m1& 6.853\,55\m3& 1.347\,10\m1& 8.759\,94\m1\\ 
37& 7.244\,80\m1& 7.284\,40\m3& 1.579\,91\p0& 3.332\,11\m4& 4.388\,22\m2& 2.537\,50\p0\\ 
  & 4.688\,90\m1& 4.758\,94\m2& 6.632\,11\m1& 6.411\,24\m3& 1.190\,07\m1& 6.379\,35\m1\\ 
38& 8.651\,51\m1& 1.460\,11\m2& 1.859\,35\p0& 6.207\,0\s\m6& 7.426\,92\m2& 2.941\,71\p0\\ 
  & 3.704\,16\m1& 4.243\,65\m2& 4.958\,50\m1& 6.025\,78\m3& 1.042\,50\m1& 4.333\,93\m1\\ 
39& 1.017\,51\p0& 2.497\,11\m2& 2.160\,00\p0& 2.163\,42\m4& 1.153\,56\m1& 3.367\,91\p0\\ 
  & 2.828\,43\m1& 3.764\,21\m2& 3.504\,70\m1& 5.689\,35\m3& 9.017\,79\m2& 2.637\,95\m1\\ 
40& 1.182\,79\p0& 3.906\,92\m2& 2.483\,81\p0& 1.154\,25\m3& 1.699\,85\m1& 3.810\,15\p0\\ 
  & 2.061\,33\m1& 3.313\,51\m2& 2.276\,54\m1& 5.388\,53\m3& 7.643\,65\m2& 1.318\,28\m1\\ 
41& 1.363\,07\p0& 5.796\,93\m2& 2.833\,14\p0& 3.151\sm\m3 & 2.417\sm\m1 & 4.242\sm\p0 \\ 
  & 1.404\,34\m1& 2.883\,01\m2& 1.286\,29\m1& 5.086\sm\m3 & 6.238\sm\m2 & 4.215\sm\m2 \\ 
42& 1.561\,51\p0& 8.335\,38\m2& 3.208\,40\p0& 6.5\sm\s\s\m3& 3.2\sm\s\s\m1& 4.5\sm\s\s\p0\\ 
  & 8.614\,88\m2& 2.460\,45\m2& 5.555\,91\m2& 4.5\sm\s\s\m3& 4.5\sm\s\s\m2& 1.9\sm\s\s\m3\\ 
43& 1.782\,63\p0& 1.175\sm\m1 & 3.592\sm\p0 & &&\\
  & 4.403\,71\m2& 2.021\sm\m2 & 1.184\sm\m2 & &&\\
44& 2.031\,89\p0& &&&&\\
  & 1.532\,30\m2& &&&&\\
45& 2.306\sm\p0 & &&&&\\
  & 1.49\s\sm\m3& &&&&\\
\end{longtable}
\end{center}

\begin{figure}[hbt]
\setlength{\unitlength}{1mm}
\begin{picture}(140,50) (-20,10) 
\put(0,0){\mbox{\scalebox{1.2}{\includegraphics{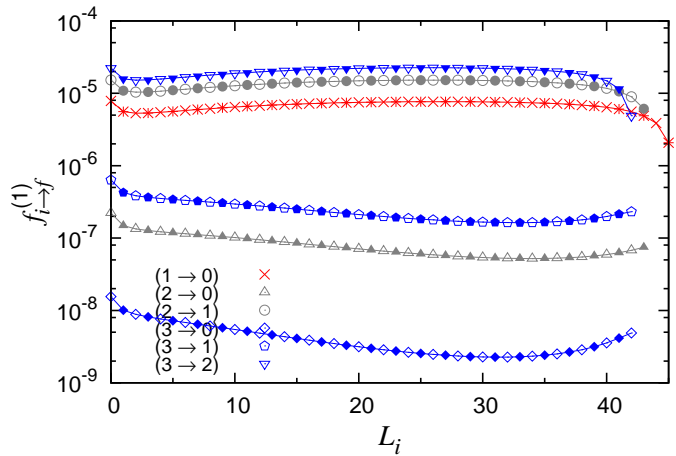}}}}
\end{picture} \\
\caption{Oscillator strengths for $L_f = L_i + 1$ transitions.}
\label{fig:2}
\end{figure}
\begin{figure}[hbt]
\setlength{\unitlength}{1mm}
\begin{picture}(140,55) (-20,10) 
\put(0,0){\mbox{\scalebox{1.2}{\includegraphics{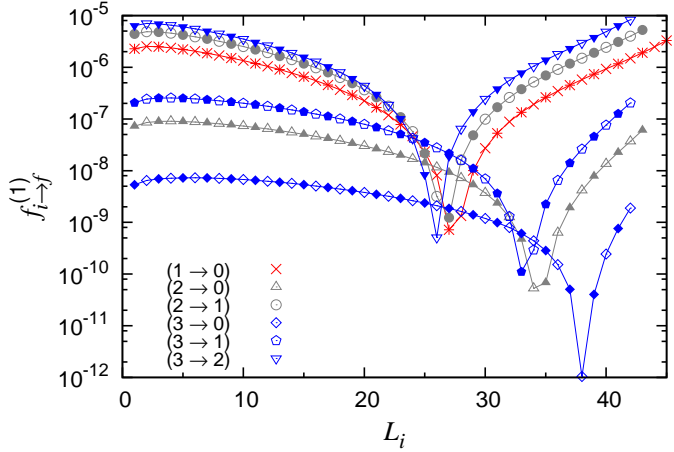}}}}
\end{picture} \\
\caption{Oscillator strengths for $L_f = L_i-1$ transitions.}
\label{fig:3}
\end{figure}

Oscillator strengths are depicted in Figs.~\ref{fig:2} and \ref{fig:3}. 
For the transitions with $\Delta L = L_i - L_f = -1$ displayed in Fig.~\ref{fig:2}, 
the oscillator strengths are smooth functions of $L$. 
The behaviour of the curves depends on $\Delta v = v_i - v_f$. 
The $\Delta v = 1$ strengths present a shallow minimum at $L_i = 2$ and a maximum 
near $L_i=26$. 
The $\Delta v = 2$ strengths are smaller by more than an order of magnitude 
and a minimum appears at $L_i = 34$ for $v_i=2 \rightarrow v_f=0$ 
and $L_i = 33$ for $v_i=3 \rightarrow v_f=1$. 
The $\Delta v = 3$ strengths are smaller by more than an order of magnitude 
than the $\Delta v = 2$ strengths and the minimum appears at $L_i = 31$. 
The presence of a minimum is due to a minimum around the same $L_i$ value 
in the matrix element appearing in \eref{2.3} 
shifted by the monotoneous decrease of the energy difference. 
The $\Delta L = 1$ strengths presented in Fig.~\ref{fig:3} display a deep minimum 
around $L_i = 27$ for $\Delta v = 1$, $L_i = 33$ for $\Delta v = 2$ and $L_i = 38$ for $\Delta v = 3$. 
These minima occur at increasing $L_i$ values with increasing $\Delta v$ 
and are all due to a change of sign of the matrix element. 

In~\cite{TTZY12}, Tian \etal not only present rovibrational energies 
for the six lowest vibrational ($v=0-5$) and rotational ($L=0-5$) levels 
but also the dipole oscillator strengths between these levels 
with the 2006 mass convention. 
Comparisons between $L_f = L_i-1$ strengths within the same rotational bands ($v_f=v_i$) 
are displayed in Table~\ref{tab:6}. 
The $L \rightarrow L+1$ strengths of \cite{TTZY12} are multiplied by $(2L+1)/(2L+3)$ 
to reverse them into $L+1 \rightarrow L$. 
The Lagrange-mesh calculations are done using the parameters $N_x = N_y = 40$, 
$N_z=20$ and $h_x = h_y = 0.08$, $h_z=0.5$. 
The agreement between both calculations improves with increasing $L_i$. 
It almost reaches 12 significant figures for $L_i = 4$, 5 and $v_i = 0$, 1. 
Using relation \eref{2.2}, dipole transition probabilities can be calculated. 
We find that the six significant figures presented in Tables~\ref{tab:4} and \ref{tab:5} are unchanged. 
These numbers are independent of later changes of mass convention for the proton and deuteron. 
\begin{table}[hbt]
\centering{
\caption{Comparison of the dipole oscillator strengths between levels of a 
same rotational band ($v_f = v_i$, $L_f = L_i - 1$). 
The mass conventions are $m_p=1836.15267247$ and $m_d=3670.4829654$. 
The Lagrange-mesh results are given in each first line. 
Each second line displays the results of Tian \etal \cite{TTZY12}.} 
\label{tab:6} 
\resizebox{16cm}{!}{
\begin{tabular}{rllll}
\hline
\hline
$L_i$&$v_i=0$&$v_i=1$&$v_i=2$&$v_i=3$\\
\hline

1&5.219\,599\,338\,877-6& 5.563\,140\,035\,05-6& 5.911\,310\,070\,3-6& 6.264\,008\,727-6\\
 &5.219\,599\,339\,113-6& 5.563\,140\,035\,53-6& 5.911\,310\,071\,0-6& 6.264\,008\,700-6\\
2&1.252\,564\,912\,793-5& 1.334\,723\,278\,42-5& 1.417\,978\,139\,8-5& 1.502\,305\,542-5\\
 &1.252\,564\,912\,744-5& 1.334\,723\,278\,42-5& 1.417\,978\,139\,7-5& 1.502\,305\,541-5\\
3&2.012\,682\,620\,839-5& 2.143\,950\,651\,96-5& 2.276\,942\,668\,5-5& 2.411\,620\,728-5\\
 &2.012\,682\,620\,915-5& 2.143\,950\,651\,96-5& 2.276\,942\,668\,4-5& 2.411\,620\,725-5\\
4&2.782\,267\,297\,866-5& 2.962\,302\,573\,95-5& 3.144\,649\,633\,9-5& 3.329\,257\,006-5\\
 &2.782\,267\,297\,892-5& 2.962\,302\,573\,95-5& 3.144\,649\,633\,2-5& 3.329\,257\,003-5\\
5&3.555\,747\,344\,258-5& 3.783\,538\,364\,46-5& 4.014\,170\,682\,2-5& 4.247\,579\,949-5\\
 &3.555\,747\,344\,241-5& 3.783\,538\,364\,45-5& 4.014\,170\,681\,9-5& 4.247\,579\,945-5\\
\hline
\end{tabular}
}}
\end{table}

\begin{figure}[hbt]
\setlength{\unitlength}{1mm}
\begin{picture}(140,60) (-20,7) 
\put(0,0){\mbox{\scalebox{1.2}{\includegraphics{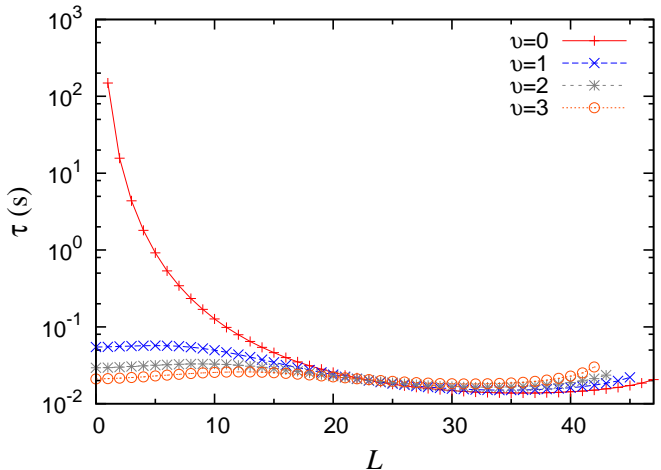}}}}
\end{picture} \\
\caption{Lifetimes $\tau$ in seconds for the first four rotational bands 
($v = 0-3$).}
\label{fig:4}
\end{figure}

For the $22$ vibrational levels of the rotationless state $L = 0$, 
lifetimes have been calculated at the Born-Oppenheimer approximation in \cite{PHB79}. 
The lifetimes $5.47 \times 10^{-2}$ s, $2.94 \times 10^{-2}$ s and $2.11 \times 10^{-2}$ s 
for the first, second and third $L = 0$ excited vibrational levels agree respectively 
with our values $5.460\,856\times 10^{-2}$ s, $2.932\,100\times 10^{-2}$ s 
and $2.102\,753\times 10^{-2}$ s, obtained from Table \ref{tab:5}. 
The lifetimes of all calculated levels are displayed in Fig.~\ref{fig:4}. 

Because of the presence of a permanent dipole moment, 
dipole transitions are dominant in the spectrum of HD$^+$ 
and the lifetimes are expected to be much smaller 
than for the quadrupole transitions in H$_2^+$ and D$_2^+$. 
Indeed, the longest lifetime is about $2.5$ min, for the $(1^-,0)$ level, 
while typical lifetimes are of the order of days for H$_2^+$ \cite{OB12T} 
and months for D$_2^+$ \cite{OP13D2}. 
Roughly speaking, except for the ground-state rotational band below $L = 13$, 
the $v \le 3$ lifetimes do not depend much on $L$ 
and their values are located between about 0.01 and 0.05 s. 
They slightly decrease with increasing vibrational excitation.  
\section{Conclusion}
\label{sec:conc}
With the Lagrange-mesh method in perimetric coordinates, 
the three-body Schr\"odinger equation of the heteronuclear molecular ion HD$^+$ 
is solved with Coulomb potentials. 
Energies and wave functions are calculated for up to four of the lowest vibrational 
bound or quasibound states from $L = 0$ to $47$. 
Lagrange-mesh results are obtained with 40 mesh points for the $x$ and $y$ coordinates 
and 14 mesh points for the $z$ coordinate.
The accuracy is around 12 digits for the lowest vibrational level 
and slowly decreases  with vibrational excitation 
until 9 digits for the third excited vibrational level. 
These accuracies are maintained along the whole bound rotational bands. 

With the corresponding wave functions, a simple calculation 
using the associated Gauss-Laguerre quadrature 
provides the electric dipole strengths and transition probabilities per time unit 
over the whole rotational bands. 
Tests with increasing numbers of mesh points and various truncations on $K$ show 
that the accuracy on these probabilities should reach at least six significant figures, 
independently of recent or future improvements in the proton and deuteron mass values. 
For low-$L$ transitions, 
the first 10 or 11 figures of our oscillator strengths agree with those of \cite{TTZY12}. 

The dipole transition probabilities of the heteronuclear HD$^+$ are much larger 
than the quadrupole transition probabilities of the homonuclear H$_2^+$ and D$_2^+$. 
Hence, lifetimes are much shorter. 
Except for the lowest levels in the ground-state rotational band $L \le 13$, 
all calculated lifetimes have an order of magnitude around $10^{-2}$ s. 
The lifetime of the first rotational excited level is about 2.5 min. 
\ack
HOP thanks FRS-FNRS (Belgium) for a postdoctoral grant.
HOP would like to thank Professor A Turbiner for his kind invitation to the Instituto de Ciencias Nucleares (UNAM, Mexico City) where the 
manuscript was completed and the financial support through the CONACyT project IN109512.
\end{document}